\documentclass[aps,prl,twocolumn,superscriptaddress]{revtex4-1}
\usepackage{graphicx}
\usepackage{graphics}
\usepackage{amsmath}
\usepackage{natbib}

\usepackage{hyperref}

\begin{document}

\title{Spatially resolved breakdown in reentrant quantum Hall states}

\author{A. V. Rossokhaty}
	\affiliation{Quantum Matter Institute, University of British Columbia, Vancouver, British Columbia, V6T1Z4, Canada}
	\affiliation{Department of Physics and Astronomy, University of British Columbia, Vancouver, British Columbia, V6T1Z4, Canada}
\author{Y. Baum}
    \affiliation{Department of Condensed Matter Physics, Weizmann Institute of Science, Rehovot 76100, Israel}
\author{J. A. Folk}
\email{jfolk@physics.ubc.ca}
	\affiliation{Quantum Matter Institute, University of British Columbia, Vancouver, British Columbia, V6T1Z4, Canada}
	\affiliation{Department of Physics and Astronomy, University of British Columbia, Vancouver, British Columbia, V6T1Z4, Canada}
\author{J. D. Watson}
	\affiliation{Department of Physics and Astronomy, and Microsoft Station Q Purdue, Purdue University, West Lafayette, Indiana, USA}
	\affiliation{Birck Nanotechnology Center, Purdue University, West Lafayette, Indiana, USA}
\author{G. C. Gardner}
	\affiliation{Birck Nanotechnology Center, Purdue University, West Lafayette, Indiana, USA}
    	\affiliation{School of Materials Engineering, Purdue University, West Lafayette, Indiana, USA}
\author{M. J. Manfra}
	\affiliation{Department of Physics and Astronomy, and Microsoft Station Q Purdue, Purdue University, West Lafayette, Indiana, USA}
	\affiliation{Birck Nanotechnology Center, Purdue University, West Lafayette, Indiana, USA}
	\affiliation{School of Electrical and Computer Engineering,  Purdue University, West Lafayette, Indiana, USA}
    	\affiliation{School of Materials Engineering, Purdue University, West Lafayette, Indiana, USA}

\date{\today}

\newcommand{\rxx}{{R}_{xx}}
\newcommand{\rxy}{{R}_{xy}}
\newcommand{\rdp}{{R}_{D}^+}
\newcommand{\rdm}{{R}_{D}^-}

\begin{abstract}
Reentrant integer quantum Hall (RIQH) states are believed to be correlated electron solid phases, though their microscopic description remains unclear.  As bias current increases, longitudinal and Hall resistivities measured for these states exhibit multiple sharp breakdown transitions, a signature unique to RIQH states.  We present spatially-resolved measurements of RIQH breakdown that indicate these breakdown signatures can be ascribed to a phase boundary between broken-down and unbroken regions, spreading chirally from source and drain contacts as a function of bias current and passing voltage probes one by one.  The chiral sense of the spreading is not set by the chirality of the edge state itself, instead depending on electron- or hole-like character of the RIQH state.
\end{abstract}

\pacs{}

\maketitle
A variety of exotic electronic states emerge in high mobility 2D electron gases (2DEGs) at very low temperature, and in a large out-of-plane magnetic field. The most robust are the integer quantum Hall states, described by discrete and highly degenerate Landau levels. When the uppermost Landau level is partially filled, electrons in that level may reassemble into a fractional quantum Hall (FQH) liquid
\cite{laughlin, halperin, fqhe} or condense into charge-ordered states, from Wigner crystals to nematic stripe phases\cite{du, lilly, pan92,koulakovprl,koulakovprb, eisenstein02, xia2pk, csathy05, goerbig, shibata}.  Such charge ordered states, or electron solids, are observed primarily above filling factor $\nu=2$, where Coulomb effects are  strong in comparison to magnetic energy scales.  They are believed to be collective in nature\cite{csathy}, prone to thermodynamic phase transitions like melting or freezing of a liquid.

Numerical simulations of electron solids indicate alternating regions of neighboring integer filling factors with dimensions of order the magnetic length\cite{Fradkin:1999bl,Spivak:2006kf}. When the last Landau level is less than half-filled, the electron solid takes the form of ``bubbles" of higher electron density in a sea of lower density (an electron-like phase).  Above half filling, the bubbles are of lower electron density giving a hole-like phase. Insulating bubble phases lead to ``reentrant" transitions of the Hall resistivity up or down to the nearest integer quantum Hall plateau, giving rise to the term ``reentrant integer quantum Hall effect" (RIQHE).

The microscopic description and thermodynamics of RIQH states remain topics of great interest\cite{smetbias, csathy,goerbig, Fradkin:1999bl,Spivak:2006kf}.  Most experimental input into these questions has come from monitoring RIQH state collapse at elevated temperature or high current bias\cite{csathy, chickering2013, Cooper:1999ji,Cooper:2003cz,smetbias, esslin}. The temperature-induced transition out of insulating RIQH states is far more abrupt that would be expected for activation of a gapped quantum Hall liquid, consistent with their collective nature. RIQH collapse at elevated temperature is apparently a  melting transition of the electronic system out of the electron solid state\cite{csathy}.

Elevated current biases also induce transitions out of the insulating RIQH state via sharp resistance steps, a phenomenon that has been interpreted in terms of sliding dynamics of depinned charge density waves\cite{Reichhardt:2005be}, or alignment of electron liquid crystal domains by the induced Hall electric field\cite{smetbias}. These interpretations assume that bias-induced phase transitions happen homogeneously across the sample. On the other hand, finite currents through a quantum Hall sample generate highly localized Joule heating. Considering the collective nature of RIQH states, this suggests a mechanism for forming inhomogeneous phases across a macroscopic sample.  

Here, we show that resistance signatures of high current breakdown for RIQH states reflect a macroscopic phase separation induced by the bias.  That is, the breakdown process itself is sharply inhomogeneous, with the electronic system after breakdown spatially fractured into regions that are either melted (conducting) or frozen (insulating).  
For all RIQH states from $\nu=2$ to $\nu=8$, the breakdown propagates clockwise or counterclockwise from the source and drain contacts with a sense that depends on the electron- or hole-like character of the particular RIQH state.  The data are explained by a phase boundary between frozen and melted regions that spreads around the chip following the location of dissipation hotspots.

\begin{figure}[t]
\includegraphics{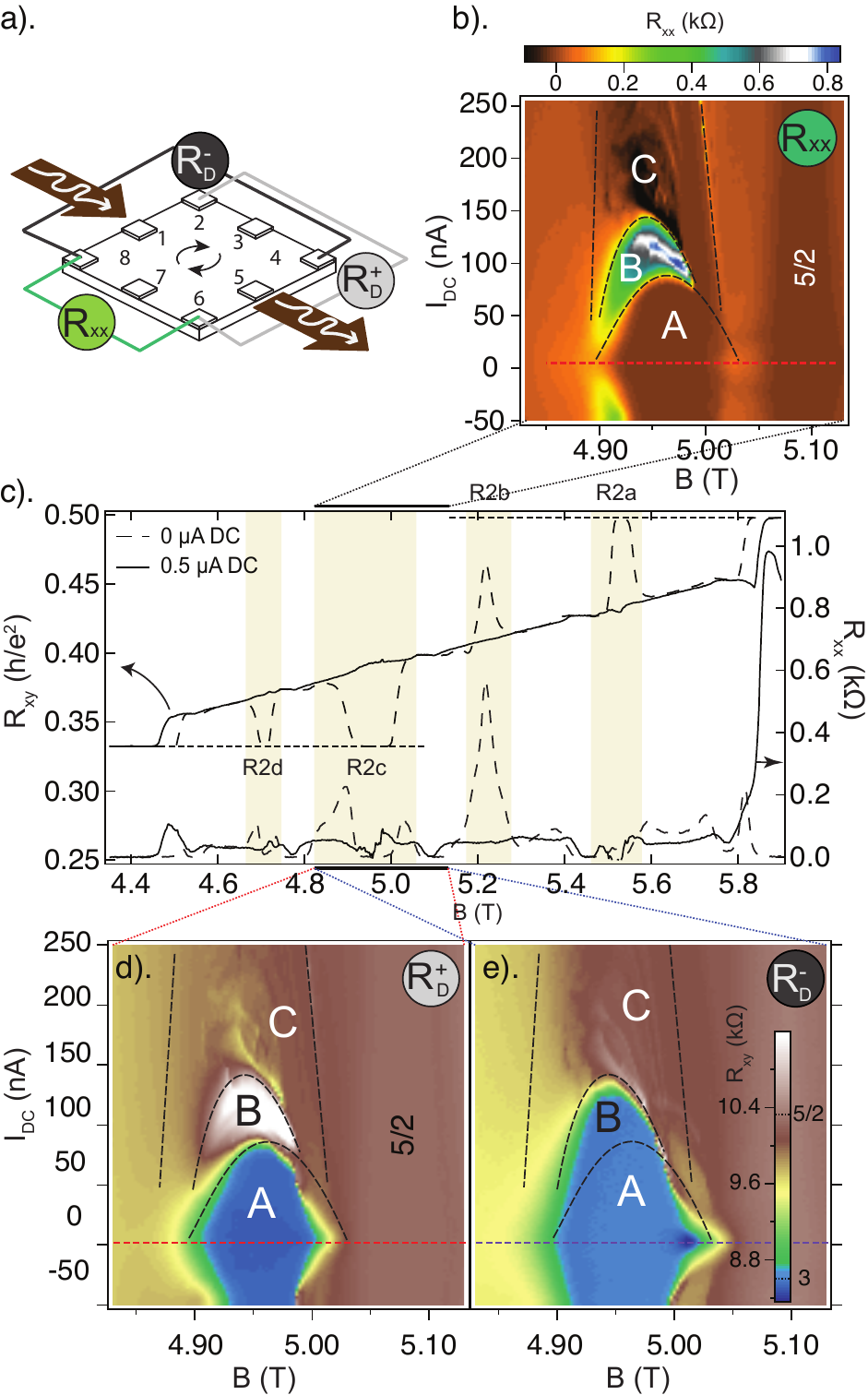}
\caption{a) Measurement schematic combining AC (wiggly arrow) and DC (solid arrow) current bias through contacts 1 and 5.
$\rxx={dV_{86}/dI}$,  $\rdp= dV_{26}/dI$,  and
$\rdm={dV_{84}/dI}$. Curved arrows indicate edge state chirality.  b) Evolution of $\rxx$ with DC bias for the  R2c reentrant and $\nu=5/2$ FQH state, showing  breakdown regions `A', `B', and `C' .  c) $\rxx$ and $\rxy$ ($dV_{37}/dI$) for filling factors $\nu=2-3$, showing the breakdown at high DC bias. (d,e) Simultaneous measurements of $\rdp$ (d) and $\rdm$ (e), taken together with data in panel b). Dashed lines are guides to the eye, denoting identical \{$B,I_{DC}$\} parameters in panels b,d,e.}
\label{4rs}
\end{figure}

Measurements were performed on a 300\,{\AA} symmetrically doped GaAs/AlGaAs quantum well with low temperature electron density $n_s=3.1\times10^{11}$~cm$^{-2}$ and mobility $15\times10^6$~cm$^2$/Vs\cite{manfrarev}. Electrical contact to the 2DEG was achieved by diffusing indium beads into the corners and sides of the 5$\times$5\,mm chip [Fig.~1a]. FQH characteristics were optimized following Ref.~\onlinecite{shiningheating}.    
Differential resistances ${R}\equiv dV/dI_b$ for various contact pairs were measured at 13\,mK by lockin amplifier with an AC current bias, $I_{AC}=5$~nA, at 71 Hz. A DC current bias $I_{DC}$ was added to the AC current in many cases.
At zero DC bias, characteristic $\rxx$ and $\rxy$ traces over $2<\nu<3$ show fragile FQH states as well as four RIQH states, labelled R2a-R2d [Fig.~\ref{4rs}c]. At high current bias the RIQH states disappear, with $\rxy$ moving close to the classical Hall resistance, while most fractional states remain well-resolved.

The RIQH breakdown process can be visualized in 2D resistance maps versus $I_{DC}$ and magnetic field.  Figure 1 presents several such maps for the hole-like R2c state ($\nu\sim 2.58$), where the Hall resistance reenters to the integer value $R_{xy}=h/3e^2$.  Breakdown transitions for $\rxx$ [Fig.~1b] divide the map into three distinct subregions  [`A', `B', `C'], similar to observations by others\cite{smetbias, esslin}.
Region A is characterized by very low $\rxx$: here the electron solid state is presumably pinned and completely insulating. The sharp transition to region B corresponds to a sudden rise in $R_{xx}$, while for higher bias (region C) the differential resistance drops again to a very small value.

The sharp transitions in the RIQH state breakdown [Figs.~1b] are entirely absent from the neighbouring $\nu=5/2$ state, a distinction seen for all RIQH states compared to all fractional states. Data for many cooldowns and RIQH states were slightly different in the details, but qualitatively consistent.  Furthermore, qualitative signatures at each pair of voltage probes ($\rxx$, $\rxy$, or the diagonal measurements $\rdp$ or $\rdm$ [Fig.~1a]) did not depend on the specific contacts used in the measurement, but only on the arrangement of the contacts with respect to source/drain current leads (see supplement).

The observation of sharp delineations in the resistance of a macroscopic sample, measured between voltage probes separated by 5\,mm, might seem to imply that the entire sample must suddenly change its electronic state for certain values of bias current and field.  Then one would expect simultaneous jumps in resistance monitored at any pair of voltage probes, albeit by differing amounts.  Comparing the three pairs of voltage probes in Figs.~1b, 1d, and 1e, one sees immediately that this is not the case.   $\rdp$ exhibits  transitions at precisely the same parameter pairs \{$B,I_{DC}$\} as $\rxx$, but for $\rdm$ no resistance change is observed at the dashed line corresponding to the $\rxx$ A-B transition.  It is well known that $\rdp$ and $\rdm$ can be different when the sample is inhomogeneous\cite{beenakkervanhouten, butikker}.  However, the extremely high quality 2DEG samples measured here are  intrinsically homogeneous, as evidenced by the visibility of closely-spaced and fragile fractional states.

\begin{figure}[t]
\includegraphics{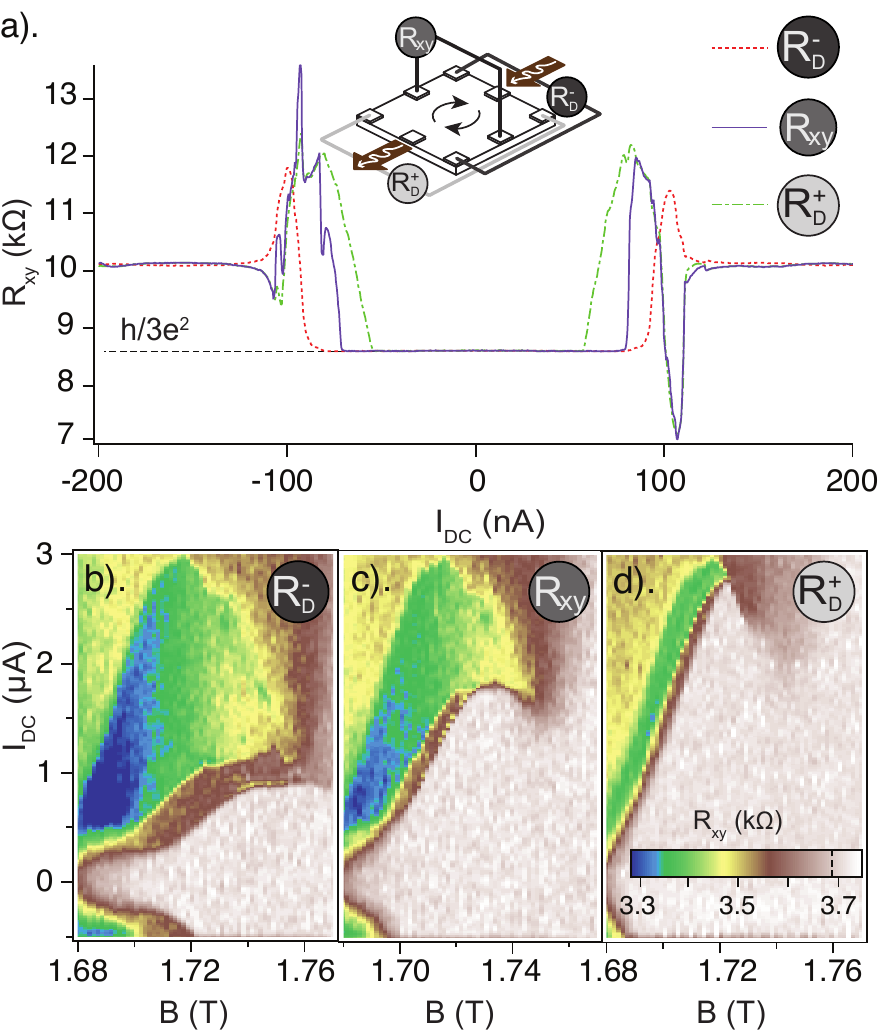}
\caption{(a) Simultaneous measurements showing the evolution of $\rdp$, $\rxy$, and $\rdm$, with DC bias, in the middle of the R2c reentrant state ($I_{AC}$=5\,nA); note that this measurement uses a contact configuration rotated by 90$^\circ$ from Fig.~1. Evolution of (b) $\rdm$, (c) $\rxy$ and (d) $\rdp$ for the R7a reentrant state with DC bias.}
\label{idep}
\end{figure}

$\rdp$ and $\rdm$ contacts are distinguished by the chirality of quantum Hall edge states: moving from source or drain contacts following the edge state chirality, one first comes to the $\rdp$ contacts, then to $\rxy$ contacts in the middle of the sample, and finally to the $\rdm$ contacts.
The bias where the A-B transition occurs for $\rdp$, $\rxy$ and $\rdm$ simply follows the spatial distribution of the respective voltage contacts,  as shown in Fig.~2a.
An analogous breakdown behaviour (breakdown bias for $\rdp$ lower than for $\rxy$, lower than for $\rdm$) was consistently observed for every hole-like RIQH state.
For all electron-like states, a similar breakdown progression was observed but the order was opposite: breakdown bias for $\rdp$ higher than for $\rxy$, higher than for $\rdm$ (see Figs.~2b-d for R7a).  


The correlation between electron/hole character and breakdown chirality offers an important hint as to the origin of this effect.  Edge state chirality is fixed by magnetic field direction, and would not suddenly reverse when crossing half-filling for each Landau level.    Instead, we propose an explanation based on localized dissipation in the quantum Hall regime, a phenomenon that is known to give rise to ``hotspots" any time a significant bias is applied to a quantum Hall sample.

Consider current injected into a sample in the integer quantum Hall (IQH) regime, where $\rho_{xx}$ is close to zero but $R_{xy}$ is large.  Driving a current $I_{b}$ through such a sample requires a potential difference $R_{xy}I_{b}$ between source and drain, and this potential drops entirely at the source and drain contacts (no voltage drop can occur within the sample since $\rho_{xx}\rightarrow 0$).  Specifically, the voltage drops where the current carried along a few-channel edge state is dumped into the metallic source/drain contact---a region of effectively infinite filling factor.

Moving now to a sample in the {\em reentrant} IQH regime, with $\rho_{xx}\rightarrow 0$ as before, hotspots again appear at any location where current flows from a region of higher to lower $R_{xy}$.  But now the local value of $R_{xy}$ is strongly temperature dependent, with a sharp melting transition in both longitudinal and transverse resistances\cite{csathy}.  The electron-like R2a reentrant state, for example, has $R_{xy}^{reentrant}=h/2e^2$ in the low temperature, low bias limit [Fig.~1c], but at higher temperature or bias the state melts to $R_{xy}^{melted}\simeq h/2.35e^2$.  In general, electron-like states have $R_{xy}^{melted}<R_{xy}^{reentrant}$ whereas hole-like states have $R_{xy}^{melted}>R_{xy}^{reentrant}$.  

For low current bias in the RIQH regime, the entire sample is effectively at integer $\nu$ and only the two hotspots associated with IQH are observed, at source and drain contacts.  As the bias increases, the regions around the two IQH hotspots melt and an extra two ``RIQH hotspots" appear.  Fig.~3 shows a numerical analysis of dissipation in a sample where the bulk is at one value of $\rxy$ (in this case, $R_{xy}^{reentrant}$) and the regions around source and drain contacts are at another ($R_{xy}^{melted}$).  RIQH hotspots appear where current passes from bulk to melted, or melted to bulk regions, depending on the relative values of $R_{xy}^{melted}$ and $R_{xy}^{reentrant}$, and are therefore different for electron-like [Fig.~3a] and hole-like [Fig.~3b] states. 
The semicircular shapes of the melted regions in Fig.~3 are defined by the simulation inputs, but in reality the melted regions would be expected to spread in the direction of extra heating, that is, following the hotspot locations, until heat flow into the substrate balances the hotspot dissipation.

\begin{figure}
\center
\includegraphics[scale=1]{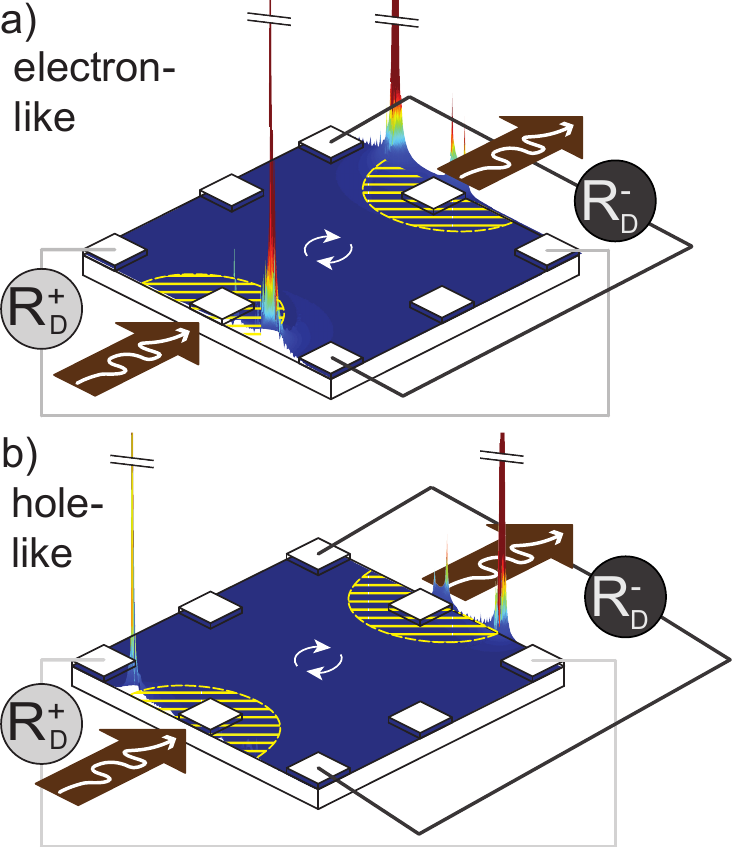}
\caption{Classical simulation of dissipation (colorscale and 3D projection) due to current flow, in a sample divided into regions with different $R_{xy}$: hatched semicircles correspond to a melted state near each contact with $R_{xy}=h/(2.5e^2)$, while the bulk (dark blue) is the reentrant state with $R_{xy}=h/(2e^2)$ (a) or $R_{xy}=h/(3e^2)$ (b).  Simulation shows hotspot locations but does not accurately capture relative magnitudes of dissipation in different hotspots.   Hotspots appear at different corners of the melted region in a) and b) (peaks extend nearly an order of magnitude higher, cut off here for clarity).}
\label{idep} 
\end{figure}

Consider as an example the R2c measurement in Fig.~1, with current from contact 1 to 5 [Fig.~1c].  RIQH hotspots for hole-like states are downstream from source/drain contacts following edge state chirality [Fig.~3b], so the melted/frozen boundary would propagate clockwise around the sample edge from contacts 1 and 5.  Within region A, we speculate that the hotspots have not yet passed a voltage probe, so no change is observed in $R_{xx}$, $R_D^+$, or $R_D^-$. When the hotspots pass voltage probes 2 and 6, used for $R_{xx}$ and $R_D^+$, both resistances register a jump due to the potential drop at the hotspot. $R_D^-$ is unaffected, because the potential drop did not pass into or out of the pair of $R_D^-$ contacts (4,8).  This mechanism also explains the progression of A$\rightarrow$B transitions for \{$R_D^+,R_{xy},R_D^-$\} in Fig.~2.  For R2c [Fig.~2a], the hotspot first passes the $R_D^+$ probe, then the $R_{xy}$ contact, then the $R_D^-$ contact; for the electron-like R7a [Fig.~2b,c,d], the hotspot propagates against the edge state chirality, so it passes the $R_D^-$ probe, then $R_{xy}$, then $R_D^+$.

\begin{figure}
\includegraphics[scale=1]{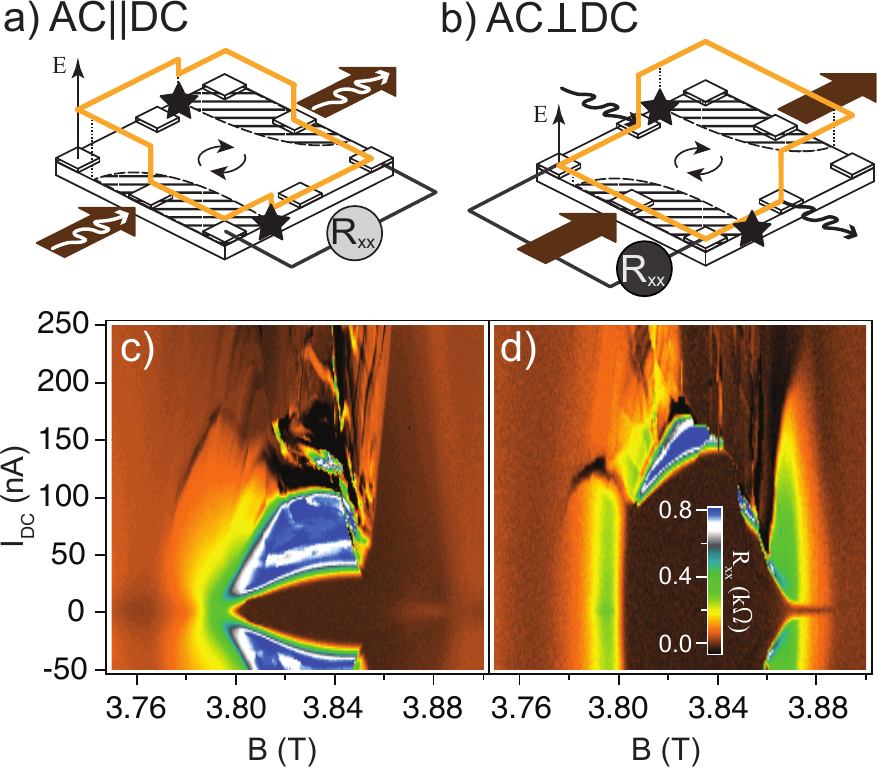}
\caption{Comparison of measurement geometries a) $AC||DC$ and b) $AC\perp{DC}$; arrows label source and drain contacts for DC (solid) and AC (wiggly) bias, and edge state chirality (curved).  Vertical axis `E' denotes local AC edge state potential (yellow line).  Hatched areas are hypothetical melted regions for an electron-like reentrant state at intermediate DC bias, $I_{DC}\sim50$nA in panels (c,d). Hotspots at the melted/frozen boundary indicated by $\star$.  (c,d) R3a $R_{xx}$ maps in $(I_{DC},B)$ plane for c) $AC||DC$ and d) $AC\perp{DC}$ measurement.}
\label{idep} 
\end{figure}

Finally, we turn to a measurement configuration that has been used to investigate possible anisotropy in the electron solid at high bias, when the Hall electric field is large.  Ref.~\onlinecite{smetbias} compared $R_{xx}$ measured parallel or perpendicular to a large DC current bias, by rotating the $R_{xx}$ voltage probes and AC current bias contacts by 90$^\circ$ with respect to the DC bias contacts [Fig.~4a,b]. It was observed that the low-$R_{xx}$ region A extended to much higher bias for the ($AC\perp{DC}$) orientation, compared to the conventional ($AC||DC$) orientation.  While Ref.~\onlinecite{smetbias} focused on R4 states exclusively, we found analogous behaviour for all reentrant states measured [see e.g Figs.~4c,d].

This behaviour can be simply explained by the hotspot-movement mechanism outlined above, without resorting to induced anisotropy in the electron solid.  Figs.~4a and 4b schematics include dashed lines to show hypothetical melted-frozen boundaries at an intermediate bias, with associated RIQH hotspots marked by $\star$'s.  The melted region surrounds the DC (not AC) current contacts, because the measurement is done in the limit of vanishing AC bias. The boundary is not symmetric around the DC contacts as the melted region is presumed to have propagated counterclockwise (for electron-like states) from the contacts, following the $\star$ hotspot locations.

The local AC potential along the edge of the sample drops sharply when passing the AC source and drain (the conventional IQH hotspots), but a second smaller potential drop occurs at each $\star$ when the melted region includes an AC source/drain [e.g.~Fig.~4a].  
For the distribution of melted and frozen phases indicated in Fig.~4, the $\star$ hotspot potential drop occurs between the $R_{xx}$ voltage probes in Fig.~4a, but not in Fig.~4b, so large $R_{xx}$ would be registered  only in the $AC||DC$ configuration. A frozen/melted configuration like that shown in Figs.~4a,b might correspond to intermediate bias, around 50 nA in Figs.~4c,d, thus explaining the large region of high $\rxx$ in Fig.~4c that appears only above 100 nA in Fig.~4d.

In conclusion, we demonstrated that bias-induced breakdown of the RIQH effect is inhomogeneous across macroscopic (mm-scale) samples. As bias increases, the RIQH breakdown propagates away from source and drain contacts with a chiral sense that depends on the electron- or hole-like character of the reentrant state, leading to different critical breakdown biases for different pairs of voltage probes.  This phenomenon may result from opposite hotspot locations for the two types of reentrant states, giving rise to melted (no longer reentrant) regions near source and drain contacts that spread in opposite directions as bias increases.  This experiment shows the danger in interpreting macroscopic measurements at a microscopic level, especially where electronic phase transitions are sharp and phase segregation may occur.

\begin{acknowledgments}
The authors acknowledge helpful discussions with J. Smet and A. Stern.  Experiments at UBC were supported by NSERC, CFI, and CIFAR.  The molecular beam epitaxy growth at Purdue is supported by the U.S. Department of Energy, Office of Basic Energy Sciences, Division of Materials Sciences and Engineering under Award DE-SC0006671. 
\end{acknowledgments}

\bibliography{main}
\clearpage
\renewcommand{\figurename}{S.}
\setcounter{figure}{0}
\section{Supplement}
\subsection{How do $\rdp/\rdm$ measurement depend on contacts?}
Figure S.\,2 contrasts diagonal measurements of reentrant states between $\nu=3-4$ for contact configurations rotated by 90$^\circ$ around the chip.  Almost all of the major characteristics of the RIQH breakdown are the same for the two sets of contacts, demonstrating that the effects described in the main text do not depend on specific contact imperfections but rather on relative location of voltage and current contacts.

\begin{figure}[h]
\includegraphics[scale=1]{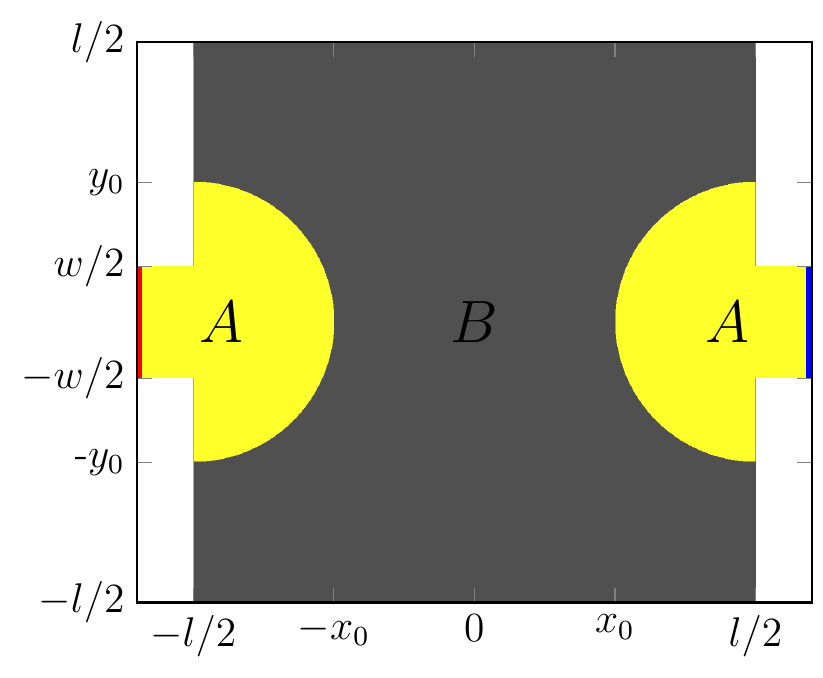}
\caption{Geometry of the domain $\Omega$, where simulation is performed. Regions A and B denote the areas with different $\sigma_{xy}$'s in the simulation.}
\end{figure}

\begin{figure*}[h]
\includegraphics[scale=0.92]{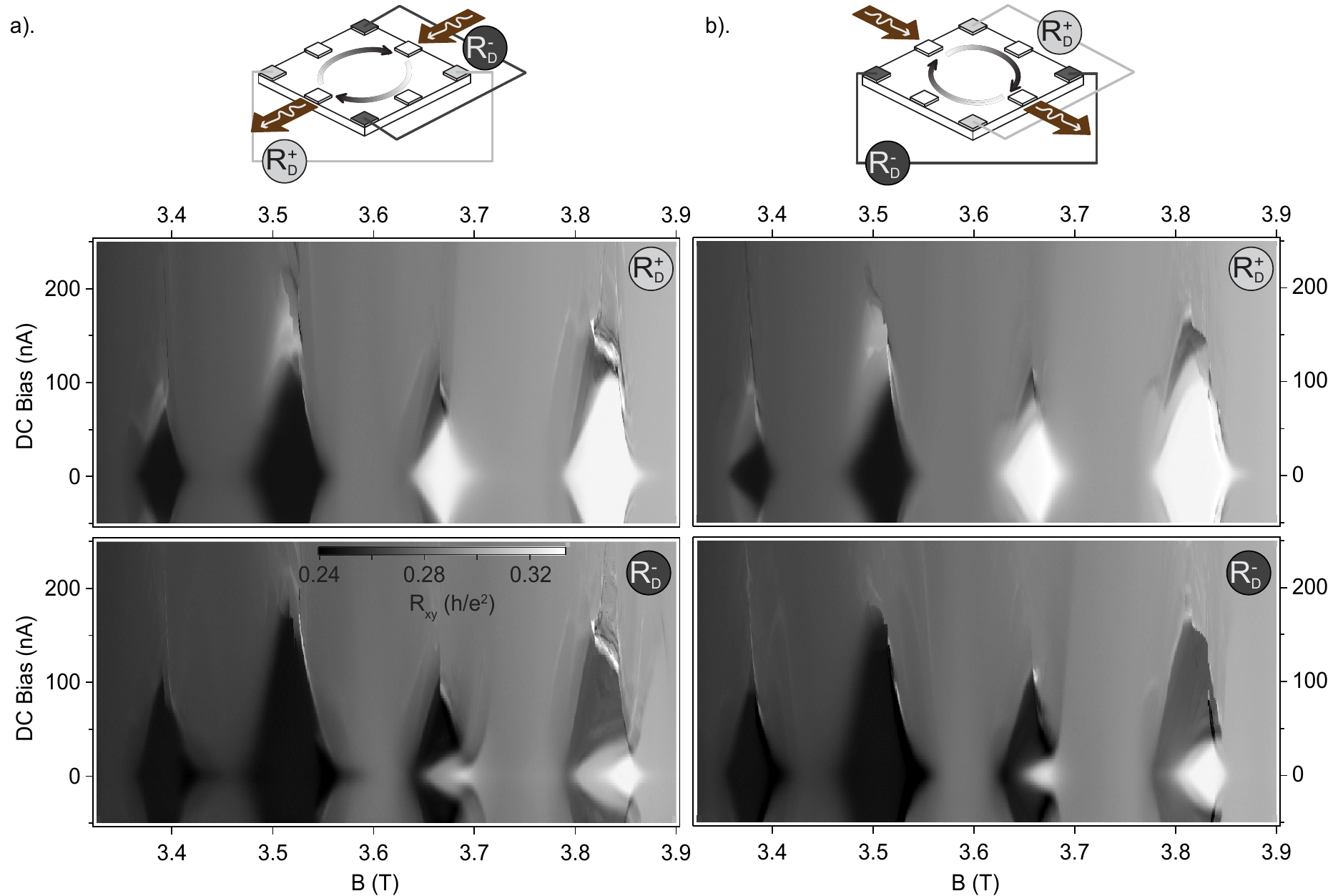}
\caption{Diagonal measurements, $\rdp$ and $\rdm$, with the current flowing in two perpendicular orientations with respect to the sample axes. Note that $\rdp$ and $\rdm$ are, as always, defined with respect to the source and drain contacts.  Most features of the data are reproduced for both orientations, indicating that they depend not on specific contacts but only on the relative location of voltage probes with respect to source and drain.  Specifically, it is clearly seen that which of the two diagonal measurements breaks down at a higher bias (for a given reentrant states) stays the same for the different current orientations.}
\end{figure*}

\subsection{Simulation details}
The results are based on a classical solution of transport equations (Kirchhoff's laws). Considering a two dimensional domain $\Omega$, for any point $(x, y)\in\Omega$, Kirchhoff's laws dictate:
\begin{equation}
\nabla\cdot{j}=0, 
\nabla\times{E}=0,
\end{equation}
where j and E are the current density and electric field respectively. Introducing the electric potential, $E=-\nabla\phi$, the equation for E is trivially fulfilled. Finally, assuming the local relation $j=\sigma{E}$, we get:
\begin{equation}
\nabla\cdot{(\sigma\nabla\phi)}=0,
\end{equation}
where $\sigma$ has the general form:
\[
\sigma=
\begin{bmatrix}
\sigma_{xx} & -\sigma_{xy}\\
\sigma_{xy} & \sigma_{xx}
\end{bmatrix}
\]

As long as $\sigma_{xx}>0$, (2) is an elliptic differential equation, and therefore, the existence of an unique solution for $\phi$ is guaranteed for any combination of boundary conditions (BC). Notice that, although we are interested in the quantum Hall effect, at the level of the semi-classical equations we cannot consider the strict limiting case $\sigma_{xx}=0$, since the equation will no longer be elliptic and hence not solvable. As we are interested in high magnetic fields, $\sigma_{xy}/\sigma_{xx}\sim 100$ is used.

Next, the geometry, $\Omega$, and the BC should be defined. The geometry is shown in S.\,1.  In our simulation we consider two cases for BC:
\begin{enumerate}
\item $\sigma_{xy}^A=2.5e^2/h$, $\sigma_{xy}^B=2e^2/h$
\item $\sigma_{xy}^A=2.5e^2/h$, $\sigma_{xy}^B=3e^2/h$
\end{enumerate}
In both cases $\sigma_{xx}=0.02e^2/h$ for any $(x, y)\in\Omega$. A current I is injected uniformly through the red boundary (contact) and it is collected from blue boundary.

The PDE in (2) is solved with a finite element method, by casting the PDE into integral forms and optimizing to the weak solution among the finite dimensional vector space of continuous piecewise linear function on a triangular grid.

\end{document}